\documentclass[aps,pra,groupedaddress,showpacs]{revtex4}
\usepackage[latin1]{inputenc}
\usepackage{amsmath}
\usepackage{amsfonts}
\usepackage{amssymb}
\usepackage[dvips]{graphicx,graphics}
\usepackage{epsfig, caption}
\usepackage{xcolor,soul}
\sethlcolor{yellow}
\setulcolor{red}
\newcommand{\beq}{\begin{equation}}
\newcommand{\eeq}{\end{equation}}

\begin{document}
\title{An ion ring in a linear multipole trap for optical frequency
metrology}
\author{C. Champenois}
\email[]{caroline.champenois@univ-provence.fr}
\author{M. Marciante}
\affiliation{Physique
des Interactions Ioniques et Mol\'eculaires, UMR 6633 CNRS et Aix-Marseille Universit\'e,
 Centre de Saint J\'er\^ome, Case C21,
13397 Marseille Cedex 20, France}
\author{M. Kajita}
\affiliation{National Institute for Information and Communications Technology, 4-2-1, Nukui-Kitamachi, Koganei, Tokyo 184-8795, Japan}
\author{J. Pedregosa-Gutierrez}
\author{M. Houssin}
\author{M. Knoop}
\affiliation{Physique
des Interactions Ioniques et Mol\'eculaires, UMR 6633 CNRS et Aix-Marseille Universit\'e,
 Centre de Saint J\'er\^ome, Case C21,
13397 Marseille Cedex 20, France}

\date{\today}

\begin{abstract}
A ring crystal of ions trapped in a linear multipole trap is studied
as a basis for an optical frequency standard. The equilibrium
conditions and cooling possibilities are discussed through an
analytical model and molecular dynamics simulations. A configuration
which reduces the frequency sensitivity to the fluctuations of the
number of trapped ions is proposed. The systematic shifts for the
electric quadrupole transition of calcium ions are evaluated for
this ring configuration. This study shows that a ring of 10 or 20
ions allows to reach a short term stability better than for a single
ion without introducing limiting long term fluctuations.

\end{abstract}


\maketitle

\section{Introduction}
\label{s-introduction} Progress in atom and ion cooling and
trapping, laser stabilization and high-resolution  spectroscopy
makes narrow optical transitions the best candidates  for frequency
standards. Three experimental  systems take the optical clocks to
their highest performances:  a laser-cooled neutral atom cloud in a
MOT \cite{udem01,degenhardt05},  an ensemble of laser-cooled neutral
atoms in  an optical lattice \cite{takamoto05,akatsuka08,lemke09}
and single trapped laser-cooled ions
\cite{margolis04,madej04,schneider05,rosenband08}. All systems have
advantages and drawbacks. The main advantage of single ion
experiments is the quasi-perfect control of the internal and
external degrees of  freedom, permitted by the trapping in
radiofrequency (rf) electric fields in the  quadrupole
configuration. This ultimate control,  possible only with a single
particle, is paid for by a low signal-to-noise ratio which requires
a long integration time, resulting in a small short term stability
when compared to neutral atom optical clocks where $10^6$ to $10^8$
atoms are interrogated at the same time.

In the same time,  microwave frequency standards with large clouds of
trapped ions ($>10^6$ ions) \cite{fisk97,prestage06} are developed
to assure a good  short term stability (characterized by an Allan deviation
$\sigma_y(\tau) < 10^{-13} /\sqrt{\tau}$) and an excellent long term stability
$\sigma_y(\tau) < 10^{-16}$ per day).  In the context of deep space navigation
\cite{prestage07}, precision is not a major issue and a fractional frequency
uncertainty of $10^{-11}$ is admitted to be sufficient. 
The main systematic effect which limits the long term stability is
the second order Doppler effect induced by the rf driven motion
(called micromotion). This effect depends on the size of the cloud
through the maximum amplitude of the electric field seen by the ions
and requires a high stability of the ion number over a long time
scale. For an equal number of trapped ions, this shift is reduced by
using linear multipole traps where the electric field amplitude
is almost null  in the center part of the trap 
\cite{prestage99}. Large ion samples can thus be trapped with
reduced micromotion, compared to the quadrupole geometry. For a
dense enough ion cloud, one can show \cite{prestage07} that the
maximum second order Doppler shift $\delta f_{D2}=-f_0 \ \langle
v^2\rangle /(2c^2)$ simply writes \beq \delta  f_{D2}=-f_0\frac{q^2
N_L}{8 \pi \epsilon_0 m(k-1)c^2}, \label{eq_D2mw} \eeq where $f_0$
is the frequency of the atomic clock transition, $N_L$ is the number
of ions per unit length, $m$ is the mass of the ions, $q$ their
charge, $2k$ is the number of electrodes in the multipole trap and
$c$ the speed of light. The $(k-1)^{-1}$ scaling justifies
the choice of a higher order multipole \cite{prestage01} for
challenging clock applications. Indeed, in \cite{prestage07} a 16-pole trap is used to trap ions during the interrogation of their  clock transition. Multipole rf traps are also widely used to study cold reactive
collisions \cite{gerlich92} as cooling with a buffer gas from 300 K
to 4 K \cite{wester09}   allows  to control the kinetic energy of
the collisions and  to reproduce astrophysical conditions.

Laser cooling and  crystal structures of cold ions in multipole
traps are far less studied than in quadrupole traps. Laser cooling
and observation of ion crystals in a linear octupole trap have been
reported in \cite{okada07,okada09}. A semi-analytical and  numerical
study of the structure, scaling laws and phase transitions of cold
ions in an isotropic 3D octupole can be found in \cite{calvo09}.
These two studies show that a cold ion cloud can be described as a
hollow core system, resulting from the balance between the Coulomb
repulsion and a trapping potential nearly flat
in the center and very steep at the border. Such a geometry  can be
also deduced by a cold fluid model \cite{champenois09}. When the
number of trapped ions is reduced to the order of 100 or less,
simulations show that, for certain trapping parameters,  the tube
formed by the ions reduces to a ring crystal centered on the
symmetry axis.

In this article, we propose   to use laser-cooled ions trapped in a
linear multipole rf trap and organized in a ring structure as a
basis for an optical clock. The aim is not to compete with the
highest precision a single ion optical clock can offer, but to
propose a trade-off where a somewhat lower precision is compensated
by a gain in short term stability offered by the interrogation of
several ions at the same time.  The main technological challenge in
the realization of a single ion frequency standard is the clock
laser which has to reach frequency stabilities of the order of 1
Hz/s. The present proposal allows to reach total clock performances
comparable to a single-ion experiment by relaxing the constraints on
the laser performance by at least one order of magnitude. The
improvement can be illustrated by the Allan variance which is used
to quantify the stability : \beq \sigma_y(\tau)=\frac{1}{\pi Q
(SNR)}\sqrt{\frac{T_c}{\tau}}, \eeq with $Q$ the quality factor of
the measured transition, $(SNR)$ the signal-to-noise ratio of the
excitation probability measurement and $T_c$ the cycle time required
to build an atomic signal to counteract on the laser frequency. The
$(SNR)$  is limited by the quantum projection noise \cite{itano93}
to  $\sqrt{N}$ and using $N=10$ or 20 ions allows to reach the same
stability value 10 or 20 times faster than  a single ion clock, if
the line is not broadened.

Compared to a chain of ions in a linear quadrupole trap, this ring
configuration  has the advantage of having the same symmetry as a
transverse laser beam intensity profile assuring that every
ion sees the same laser intensity. The second advantage of this
symmetry is that the motion along the trap axis is characterized by
a single oscillation frequency. Furthermore, the trapping parameters
can be chosen to constrain the radial size of the ring, independent
from the number of trapped ions, to first order (if this number
obeys some stability conditions). As a consequence, the possible
loss of one ion during long time operation induces only a second
order perturbation on the ring equilibrium radius and
causes very small frequency fluctuations. The compactness
and  the symmetry of a ring compared to a chain made of the same
number of ions is another advantage regarding the dispersion of the
systematic shifts induced by local electric or magnetic fields.

The planar crystal structure of the present study could also be of
use for quantum computation experiments. Planar crystals in a
quadrupole trap are proposed in \cite{porras06} as an appropriate
system for large scale quantum computation and their structure is
studied in \cite{buluta08} by molecular dynamics simulations. The ring 
configuration we propose here for trapped ions could also be used for quantum simulations like demonstrated  in \cite{olmos09}
with several Rydberg atoms at each site, to investigate many-body quantum states.

In the following section, we start by describing the features
concerning  the trapping and cooling of the sample, relevant for an
optical clock of this kind. We then identify two configurations that
obey the clock operation conditions: a ring of 10 ions and one of 20
ions.  In section \ref{s_rfshifts}, the scaling laws and numerical
values of the systematic shifts due to the ring configuration are
given. In a second step (section \ref{s_others}), the systematic
shifts usually encountered by a  single ion clock are estimated for
this ring configuration.  In the last part, all contributions are
summarized to propose an uncertainty budget for the 10- and for the
20-ions ring clock. The numerical calculations quantifying all the
effects use the optical quadrupole transition of $^{40}$Ca$^+$ as
the clock transition \cite{champenois04, kajita05} and are carried
out for a linear octupole trap ($2k=8$). However, most of the
equations can be transposed to other species and to higher order
traps.

\section{Self-organization of ions in a ring}\label{s_ring}
\begin{figure}
\begin{center}
 \includegraphics[height=6.cm]{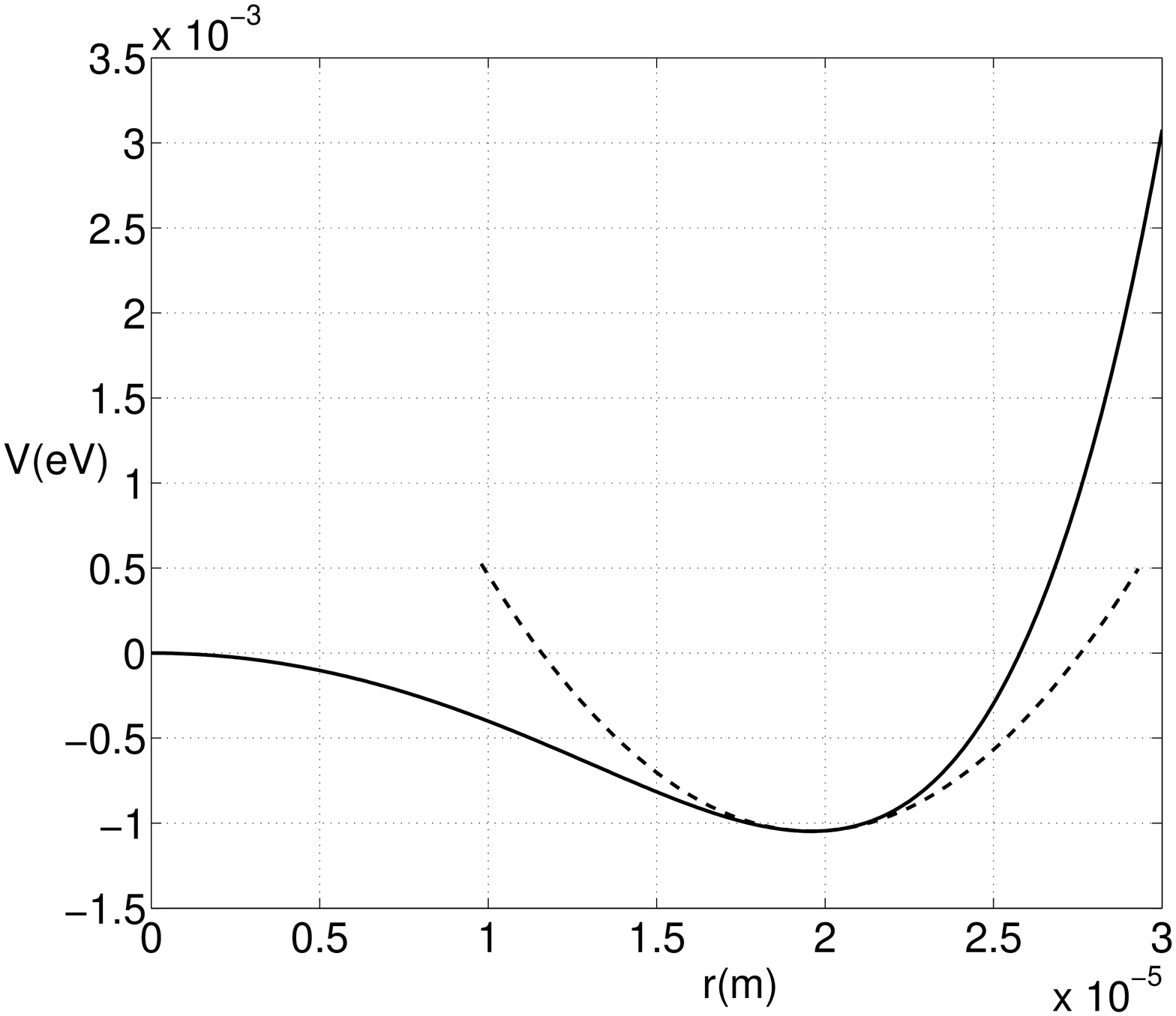}
 \end{center}
 \caption{Solid line: effective static potential (in eV) in the radial plane of an octupole trap ($2k=8$), resulting from the contributions of $V^*(r), V_{stat}$. Dashed line: the harmonic approximation of the potential well defined by $r_{min}$ (Eq.~\ref{eq_rmin}) and $\omega_{eff}$~(Eq.~\ref{eq_weff}). The potentials are defined by $\omega_z/2\pi=1$ MHz, $\Omega_z/2\pi=20$ MHz, $V_0=394.4$ V and $r_0=200 \ \mu$m, for calcium ions ($m=40$ a.m.u).  }
  \label{fig_pot}

 \end{figure}

In the adiabatic approximation, the static pseudopotential
associated to the rf electric field allows to explain the dynamics
and equilibrium positions of the ions in the transverse plane of the
trap. For a perfect linear multipole trap of order $2k$, it is
defined like \cite{gerlich92} \beq V^*(r)= \frac{k^2 q^2 V_0^2}{16 m
\Omega^2 r_0^2}\left(\frac{r}{r_0}\right)^{2k-2}, \label{eq_pseudo}
\eeq where $r$ is the ion distance to the trap axis, $V_0$ is the
amplitude of the rf potential difference between two neighbouring
electrodes, $\Omega/2\pi$ the frequency of this applied voltage and
$r_0$ the inner radius of the trap. To effectively trap  the ions, a
static potential has to be applied on electrodes placed on both ends
of the linear trap. This geometry creates a potential well along the
axis of symmetry $Oz$ which we can consider as harmonic in the centre of
the trap. The full contribution of this static potential can  be
written like \beq V_{stat}=\frac{1}{2}m\omega_z^2
\left(z^2-\frac{r^2}{2}\right) \label{eq_static} \eeq where the
harmonic well is characterized by the oscillation frequency
$\omega_z$. When the multipole is not a quadrupole ($2k \neq 4$),
the two contributions $V^*(r)$ and  $V_{stat}$ for the trapping potential
result in a potential shape (solid line in figure \ref{fig_pot}),
where the minimum is shifted from $r=0$ to $r=r_{min}$ defined by
\beq r_{min}^{2k-4}=\frac{1}{k-1}\left(\frac{2 m \Omega \omega_z
r_0^{k} }{ k q V_0}\right)^2. \label{eq_rmin} \eeq

\subsection{Equilibrium}\label{s_equilibrium}
The equilibrium position of a set of ions  results from the balance
between the Coulomb repulsion and the trapping
potential. An analysis of the forces along the radial direction
shows that two regimes of organization can be defined, based on the
strength of the deconfinement contribution $-m\omega_z^2 r^2/4$
compared to the Coulomb repulsion induced on an ion at radial
distance $R$ by all the other ions. In the assumption of a ring of
radius $R$ in a single plane,  the Coulomb potential energy for a
set of $N$ ions $E_p^C(N)$  is
  \beq
E_p^C(N)=\frac{q^2}{4\pi \epsilon_0}\frac{N}{2}\frac{S_1(N)}{2R}
\label{eq_potCoulomb} \eeq with $S_1(N)=\sum_{n=1}^{N-1} 1/\sin(\pi
n/N)$. If this contribution is negligible compared to the
deconfinement,  the radius of equilibrium $R$ does not depend on the
number of ions $N$.  In the prospect of a clock where the
major systematic shifts depend on the equilibrium position $R$
through the local rf electric field, this condition has to be
fulfilled to allow the ion number  not  to be
strictly reproduced from day to day. In the following, we consider
that the configuration is such that, at first order of
approximation, the equilibrium position $R$  is given by the
minimum potential $r_{min}$ defined by Eq.~\ref{eq_rmin} and does
not depend on the number of ions in the ring.  The Coulomb repulsion
can be treated as a perturbation and the radial position shift
$\epsilon$ that it induces depends on various trap parameters like
\beq
\epsilon=\frac{q^2}{4\pi\epsilon_0}\frac{ S_1(N)}{4(k-2)r_{min}^2 m \omega_z^2}.  
\label{eq_epsilon} \eeq With the ion number $N$ and ring radius
considered in the following,  the relative Coulomb shift
$\epsilon/r_{min}$ is smaller than 0.02, justifying the assumption
that $R \simeq r_{min}$ to  deduce the scaling laws for the
systematic shifts.

With the typical trapping parameters used through this article, local  potential depths  equivalent  to 10~K
can be made while temperatures as low as 10 mK can be reached in the
radial direction by Doppler laser cooling. Laser cooled ions will
settle in the bottom of this well and at this level, the radial
pseudopotential can be approximated by a  harmonic potential
centered on $r_{min}$ (see Figure \ref{fig_pot}). An analytical
analysis shows that the second order expansion of the radial
potential around $r_{min}$, $V^{*}(r)+V_{stat}(r) \simeq
V^{*}(r_{min})+V_{stat}(r_{min})+m \omega_{eff}^2 (r-r_{min})^2/2$,
is characterized by an effective oscillation frequency
$\omega_{eff}$ which depends only on the axial frequency and the
order of the multipole by
 \beq
  \omega_{eff}=\sqrt{k-2} \  \omega_z.
  \label{eq_weff}
  \eeq
This shows that the strength of the axial trapping
 $\omega_z$ also defines the strength of the radial
trapping around the equilibrium position $R$.

To  make relevant predictions based on the pseudopotential
(Eq~\ref{eq_pseudo}), the adiabatic approximation at
position $r_{min}$ has to be satisfied. Contrary to quadrupole
traps, there is no absolute criterion for  adiabatic operation of
the multipole trap. However,  the Mathieu parameter $q_x$ used in
quadrupole traps to characterize the trajectories of ions can be
generalized to multipole with the main difference that this new
parameter $\eta$ depends  on the ion's location in the trap
\cite{teloy74, gerlich92}.  For a perfect multipole, this parameter
is defined like \beq \eta(r)=k(k-1)\frac{q V_0}{m \Omega^2
r_0^2}\left(\frac{r}{r_0}\right)^{k-2} \label{eq_eta_r} \eeq and has
been empirically limited to 0.3 to guarantee stability of the
trajectories \cite{gerlich92}. A more recent experimental study
of the loss mechanism in a 22-pole trap \cite{mikosch08}
complementary to a model of effective trapping volume
\cite{mikosch07} has demonstrated stability up to $\eta(r)<0.36 \pm
0.02$. Moreover, clock operation requires  also a small
micromotion amplitude to limit all the systematic effects induced by
this motion and to minimize the rf heating that may occur, as in
quadrupole traps. For this reason, and by analogy with linear
quadrupole traps, we limit the set of trapping parameters to keep
$\eta<0.2$. For a ring of ions at a distance $R=r_{min}$ from the
center and including Eq.~\ref{eq_rmin} into Eq.~\ref{eq_eta_r}, the
adiabatic parameter takes the simple form \beq \eta(R)=2\sqrt{k-1}
\frac{\omega_z}{\Omega} \label{eq_eta} \eeq which does not
explicitly depend on the equilibrium position nor on the number of
ions, and immediately fixes the  range of the trapping frequency
$\Omega/2\pi$ once the axial oscillation frequency is chosen.
Indeed, in an octupole trap with  $\omega_z/2 \pi=1$ MHz, it takes
$\Omega/2\pi
>17$ MHz to assure $\eta<0.2$. Furthermore, the amplitude of the
micromotion $\delta R_{\mu}$ also scales like $\eta(R)$ but
increases with the distance to the center like \beq \delta R_{\mu}=R
\eta(R) \frac{1}{2(k-1)}. \label{eq_mumvt} \eeq

\begin{figure}
\begin{center}
 \includegraphics[height=5.5cm]{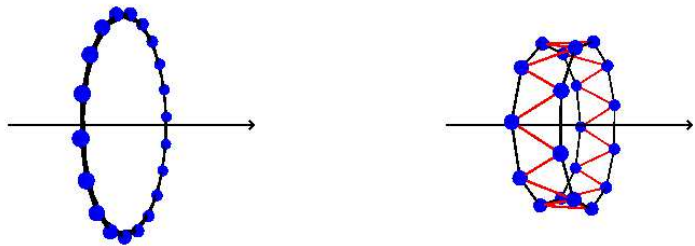}
 \end{center}
 \caption{ (Color online) Equilibrium position of 20 calcium ions around the trap symmetry axis,
calculated by MD simulations in the pseudo-potential of a linear
octupole trap. The potential is defined by $\omega_z/2\pi$= 1 MHz
and $\Omega_z/2\pi$ = 20 MHz and for 20 ions, the stability limit is
$R_l = 23 \mu$m. Left: $V_0$ = 3142 V and $r_0 $= 400 $\mu$m,
resulting in $r_{min}> R_l$, the one ring configuration is stable (R
= 28 $\mu$m). Right: $V_0$ = 5771 V and $r_0 = 400 \mu$m, resulting
in $r_{min} < R_l$, the stable configuration is a double ring analog
to a closed zig-zag chain. Here the separation between the two
planes is of the order of 3 $\mu$m and their radius is 21 $\mu$m. }
  \label{fig_ring}
 \end{figure}

Therefore, a radial equilibrium position as small as possible  is
chosen, to reduce the rf induced heating as well as the rf induced
systematic shifts, which depend on the local rf electric field and
scale like $R^{k-1}$. The lower limit of this radius is set by  the
ion-ion distance. For a distance too small, the one ring
configuration is not stable and the system relaxes to a double ring
where the sum of the Coulomb and trapping
potential energies is lower. Simple energy minimization, confirmed
by molecular dynamics (MD) simulations, shows that in a double ring
of an even number of ions, the ions alternate from one ring to the
other in a configuration analog to the zig-zag chain already
observed in a quadrupole linear trap \cite{raizen92} (see
Figure~\ref{fig_ring}). Writing the equilibrium condition along $Oz$
for a double ring of $N$ ions requires that $R$ should be smaller
than a limit $R_l$, which sets the lower stability limit for a
one-ring configuration. This limit size depends only on the strength
of the axial potential and the number of ions as \beq
R_l=\left(\frac{q^2/4\pi\epsilon_0}{4 m
\omega_z^2}\right)^{1/3}\left(\sum_{n=1}^{N/2}\frac{1}{\sin^3
\frac{(2n-1)\pi}{N}}\right)^{1/3} \simeq
\left(\frac{q^2/4\pi\epsilon_0}{2 m \omega_z^2}\right)^{1/3}
\frac{N}{\pi}. \label{eq_Rl} \eeq and therefore scales like $N$.

\subsection{Doppler cooling}\label{s_cooling}
So far, we have been concerned by the static properties of the ring
configuration. In this second section, we focus on the thermodynamic
issues related to ion temperature required for  clock operation. We
first introduce these requirements and their consequences for the
geometry of the set-up and then propose two configurations that
fulfill these requirements.

When the trapping parameters are adjusted to obtain a single ring of ions lying in the $z=0$ plane,
it seems mandatory to propagate the clock laser along the $Oz$ direction, such that no line broadening
is introduced by dispersion of the excitation strength on the clock transition from one ion to another.
In this configuration, there is no micro-motion along the laser axis and the first order Doppler effect
is sensitive only to the thermal motion along the axial direction (called macro-motion). To compete with
the existing clocks, this effect has to be canceled. The method used for trapped ion frequency standards
is to set the ion motion in the Lamb-Dicke regime where the phase modulation induced by the ion motion
inside the clock laser wave is small compared to $2\pi$. 
For optical transitions, the wavelength is so small that laser
cooling is required to reach that stage. One of the major issues of
laser cooled ion trap design is to make sure that the Lamb-Dicke
regime can be reached by Doppler laser cooling so that no other
laser cooling process is required for clock operation
\cite{itano82}.

When the ion motion can be described by an harmonic oscillation
$Z\cos \omega_z t$,  a classical expansion of the laser-atom
interaction  \cite{wineland79} shows that the Doppler spectrum can
be understood like the superposition of sidebands at frequency $f_0
\pm n \omega_z/2\pi$ with amplitude proportional to $|J_n^2(k_LZ)|$,
if the transitions are not saturated. $J_n$ is the Bessel function
of order $n$ and the modulation index $k_LZ$  depends on the clock
laser wavevector $k_L$ and the amplitude of oscillation
$Z=V_Z/\omega_z$. The velocity amplitude of the oscillation $V_Z$
depends on the temperature reached by the cooling $T_z$. When the
Doppler laser cooling limit is reached, $V_Z=\sqrt{\hbar \gamma/m}$
where $\gamma$ is the spontaneous decay rate on the cooling
transition (at 397 nm for calcium ion). For the Ca$^+$ clock
transition at 729 nm with ions at the Doppler limit $T_D$(
$T_D=0.54$ mK, with $1/\gamma=7\times 10^{-9}$ s), the modulation
index is $2 \pi \times 6.3 \times 10^5/\omega_z$. Choosing $\omega_z/2 \pi =1$ MHz is
sufficient to assure a modulation index smaller than 1 which sets
the motion in the Lamb-Dicke regime and leads to a central band
nearly 10 times stronger than the first oscillation sidebands, which
is a good enough operation condition for an optical clock\cite{champenois04}. We will
see later in the text that further increase of the axial oscillation
frequency is detrimental to the performances of the clock and
the chosen value appears to be a  good compromise between a
small oscillation amplitude and small systematic shifts.

All the  considerations introduced in the previous part about the
equilibrium  position of the ions inside the pseudopotential are not
sufficient to control the dynamics of the  set of ions. To get more
insight in the system, molecular dynamics (MD) is used to simulate
the ion motion under Coulomb repulsion  and  laser cooling, in the
full rf potential \beq \Phi_k(\mathbf{r},t)=V_0/2 \cos(\Omega t)
(r/r_0)^k \cos (k\alpha) \label{eq_phi} \eeq
where $(r,\alpha)$ are the polar coordinates in the $(x,y)$ plane. Details of these simulations will be  published elsewhere \cite{marciante_tocome}. 
These simulations show that in the single ring configuration, the
axial and radial degrees of freedom are very weakly coupled and the
thermal equilibrium of each degree is characterized by very
different temperatures. With  laser cooling on the three
directions of space,  the simulations show that for a  decoupled
system, the Doppler limit temperature is reached in the axial
direction and  the  temperatures in the radial plane  $(T_r,
T_{\alpha})$ reach 10 mK in the best configuration
(laser detuning set at $-\gamma/2$ and coupling strength defined by
a Rabi frequency equal to $\gamma/2$).  The MD simulations show that
if the trapping parameters result in a radius of the ring too close
to the stability limit $R_l$, the axial and radial degrees of
freedom are coupled and the temperature for the axial motion does
not reach the Doppler limit. Reaching the Doppler limit in
the axial direction is mandatory for the Lamb-Dicke regime, and we
can design  potentials able to trap  a ring made of 10 to 100
ions in a single plane, fulfilling this condition.

For  clock operation additional conditions have to be
fulfilled. Indeed, to prevent  light-shift on the clock transition,
the cooling lasers must   not be applied onto the ions during the
clock transition excitation \cite{champenois04}.  During this
excitation, which needs to last at least several milliseconds, the
system must remain at the same temperature.   With simulations, we
have found trapping and cooling conditions compatible with these
requirements only for rings smaller than 20 ions. The 100-ions ring
heats up very fast and, in the best configuration,  the
axial temperature increases from 0.54 mK (the Doppler limit) to 2 mK
in 1~ms. This rapid increase is due to rf heating as the micromotion
amplitude increases with $R$ (see Eq.~(\ref{eq_mumvt})) which
increases like $N$. On the contrary, for 10 ions set in a ring of
radius 20 $\mu$m or for 20 ions set in a ring of radius 40 $\mu$m,
the cooling laser can be switched off for several ms with very
little heating. To reach good statistics we build a
sequence of 2 ms of dark time (the clock interrogation)
followed by 2 ms of laser cooling, and measure the temperature
throughout the dark times. For the 10-ions ring, after 120 periods
of cooling/non cooling time, the average axial temperature is 0.47
mK and the FWHM of the distribution is 0.32 mK. Clock
operation conditions for a 10-ions ring  can thus be fulfilled
(similar results can be obtained for the 20 ion ring). For
comparison, all the configurations have the same axial and rf
frequency $\omega_z/2\pi=1$ MHz, $\Omega/2\pi=20$ MHz, and therefore
the same adiabatic coefficient at the position of the ions
$\eta(R)=0.17$. The ring radius was controlled by the
amplitude of the rf electric field.

The constraints on the trapping parameters and their relation can be
summarized as follows:  the minimum value of $\omega_z$ is fixed by
the Lamb-Dicke regime. Once $\omega_z$ chosen, the value of
$\Omega$ is fixed by the adiabacity criteria (Eq.~\ref{eq_eta}).
Then, for a given number of trapped ions $N$, the ratio $V_0/r_0^4$
is chosen to match $r_{min}$ with the smallest $R$ that allows to
keep the ions in a single ring ($R>R_l$) which does not heat up when
the cooling lasers are off. The MD analysis shows that it is
possible to form a stable ring of 10 to 20 ions, with a temperature
close to the Doppler limit in the axial direction and of the order
of 10 mK in the radial direction. In the second part of this paper
we evaluate frequency shifts on the clock transition for two
configurations tested by MD simulations : a 10-ions ring of radius
20 $\mu$m and  a 20-ions ring of radius 40 $\mu$m.  With the chosen trapping
frequencies, it takes $V_0=394$ V for $r_0=$ 200 $\mu$m to reach
$R=20 \mu$m and $V_0=1578$ V for $r_0=$ 400 $\mu$m to reach $R=40
\mu$m. Furthermore, we assume  an axial temperature of 0.54 mK (the
Doppler limit $T_D$ for calcium ion) and a radial temperature $T_r$
of 10 mK. This temperature controls the thermal motion amplitude
along the radial direction $\delta R$ through \beq
\delta R=\sqrt{\frac{2 k_B T_r}{m(k-2)\omega_z^2}  }  
\eeq
and is responsible for line broadening through every systematic shift depending on the radial position $R$.  To give an order of magnitude of this effect, for calcium ions in a linear octupole trap with $\omega_z/2\pi$=1 MHz, the thermal motion amplitude in the radial direction $\delta R$ is 0.23 $\mu$m, to be compared to the amplitude of the micro-motion $\delta R_{\mu}$ which is 0.56 $\mu$m for a 20 $\mu$m radius (and 1.1 $\mu$m for a 40 $\mu$m radius). These equilibrium and dynamic properties control several systematic shifts which are discussed in the following section.

\section{Systematic shifts induced by the rf electric field}\label{s_rfshifts}
Compared to a single ion experiment, additional systematic shifts
are introduced  by the fact that the rf electric field is not null
at the ion's equilibrium position. In this section, we focus on
these extra effects and highlight their scaling law to point out the
best compromise for a clock operation. The performances of a clock
are evaluated by its accuracy and stability and the laws scaling
their dependance with the number of ions are different. For every
shift depending on the equilibrium position $R$ (which scales with
$N$ to obey the equilibrium condition $R>R_l$), there is a line
broadening induced by the thermal oscillation amplitude $\delta R$
(which does not depend on $N$) and a long term fluctuation induced
by the possible loss of one ion. This last effect depends on $\Delta
\epsilon=\epsilon(N)-\epsilon(N-1)$ which scales  with $1/N^2$.  All
shifts are evaluated for the quadrupole transition at 729 nm
(S$_{1/2}, M_J \rightarrow$ D$_{5/2}, M_J^{'}$) of a $^{40}$Ca$^+$
ion which has no hyperfine structure. The trapping frequency is chosen to be $\Omega/2\pi = 20$~MHz, with an axial confinement
of $\omega_z/2\pi = 1$~MHz.

\subsection{Micromotion induced second order Doppler effect}\label{s_D2}
The  second order Doppler shift induced by the micro-motion is known
to be the limiting factor for microwave  ion clocks in multipole
traps.     The velocity amplitude of the micromotion $V_{\mu}$  is
equal to $\Omega \delta R_{\mu}$ and the shift induced on the clock
transition frequency $f_0$,  can be expressed like \beq \delta
f_{D2}=-f_0 \frac{\omega_z^2 R^2}{4(k-1)c^2}. \label{eq_D2} \eeq In
analogy to large clouds in a multipole trap (Eq.~(\ref{eq_D2mw})),
this shift scales with $1/(k-1)$, pleading for higher order
multipole traps. But contrary to  large clouds, the shift does not
depend explicitly on the number of trapped ions but only on the
trapping parameters, which can be very well controlled. The
compromise to make for a clock operation is clearly visible from
Eq.~(\ref{eq_D2}) :   the reduction of the second order Doppler
shift asks for $\omega_z$ and $R$ values as small as possible
whereas the ion ring stability and cooling set a lower limit to these two parameters.

For a ring of radius 20$\mu$m  (respectively 40$\mu$m) in an
octupole trap ($k=4$), the shift is $1.46 \times 10^{-14} \times
f_0=6.0$ Hz (respectively $5.85 \times 10^{-14} \times f_0=24.1$ Hz)
on the 729~nm optical clock transition of Ca$^+$. The shift itself
does not reduce the precision and stability of the clock, but its
uncertainty and fluctuations can.

The slow (on the rf time scale) thermal oscillation of the ions in
the radial plane  is responsible for a dispersion in the radial
position $\pm \delta R$ and implies a broadening of the second order
Doppler shift $\Delta \delta f_{D2}=\pm \delta f_{D2} 2\delta R /R$
(full width), which scales like $\omega_z R$, which itself is
proportional to $N$ and is lower than one Hertz with the typical
parameters we have chosen ($\Delta \delta f_{D2}=\pm 3.4 \times
10^{-16} \times f_0=\pm 0.14$ Hz for $R=20 \mu$m and $\pm 6.7 \times
10^{-16} \times f_0=\pm 0.28$ Hz for $R=40 \mu$m).

In long term operation, an ion can be lost and the shift in the
radial equilibrium position $\Delta \epsilon$ induces a modification
of the frequency shift equal to $\delta f_{D2} 2\Delta \epsilon /R$,
which scales like $1/N$. For the 10 ion ring of radius   20$\mu$m,
the fractional fluctuation is $8.5 \times 10^{-17} \times f_0= 0.03$
Hz and  it is $5.1 \times 10^{-17} \times f_0$  for a 20 ion ring of
radius   40$\mu$m. One can see that the $10^{-16}$ threshold we aim
at for the long term stability can already be reached with
a ring configuration of 10 ions.  Increasing the number of ions
decreases the long term fluctuations.

\subsection{Stark effect induced by the  electric trapping fields}\label{s_stark}
The Stark effect results from the contribution from a scalar shift,
which is proportional  to the square amplitude of the electric field
$E_{rf}^2/2+E_{dc}^2$ (the 1/2 being due to the time
averaging  of the rf component) and a contribution from a tensorial
part, which is sensitive to the angle between the magnetic field and
the electric field at the position of the ion \cite{itano00}. The
scalar and tensorial shifts are proportional to the differential
polarizability between the upper and lower  states  of the clock
transition and the scalar contribution $\Delta \alpha^0=-1.1 \times
10^{-6}$ Hz/(V/m)$^2$ \cite{arora07} is independent of their Zeeman
sublevels.  For calcium ions, the tensorial contribution is of the
same order of magnitude than the scalar one
\cite{champenois04,kajita05,arora07} and depends on the Zeeman
sublevels $M_J$ chosen for the D$_{5/2}$ state (the S$_{1/2}$ state
has no  contribution to the tensorial shift): $\Delta \alpha^2=-6.1
\times 10^{-7} \times f(M_J)$ Hz/(V/m)$^2$ \cite{arora07} with
$f(M_J=\pm 1/2)=-4/5$,  $f(M_J=\pm 3/2)=-1/5$ and $f(M_J=\pm
5/2)=1$.

The amplitude of the rf electric field is deduced from
Eq.~(\ref{eq_phi}) and for ions at distance $R$ from the center,
matching the potential minimum $r_{min}$, its  contribution to the
scalar Stark shift can be written like \beq \delta f_S^0 (rf) =
-\frac{1}{2}\Delta \alpha^0 \frac{m^2}{2(k-1)q^2} \omega_z^2
\Omega^2 R^2. \label{eq_Stark0} \eeq The scalar Stark shift induced
by the dc electric field can be expressed with respect to the rf
contribution :
 \beq \delta f_S^0(dc)= \delta f_S^0(rf)\frac{(k-1)
\omega_z^2}{2 \Omega^2}= \delta f_S^0(rf)\frac{\eta^2(R)}{8}. 
\eeq
This relation
clearly shows that for $\eta(R)\leq 0.2$, the contribution from the
dc field is negligible compared to the rf field contribution. As
the Coulomb repulsion is small compared to the dc radial electric
field, the Stark effect induced by the Coulomb interaction is
negligible compared to the dc contribution, which is itself
negligible compared to the rf contribution. Therefore, only the rf
contribution is taken into consideration in the uncertainty budget.

Like the second order Doppler shift  (Eq.~\ref{eq_D2}), the Stark
effect scales with $1/(k-1)$ and is proportional to $\omega_z^2
R^2$. The new parameters that are  introduced in the shift
dependence are the mass of the ion and the trapping frequency.
With the same  parameters as mentioned before,   $\delta f_S^0 (rf)
= 1.0 \times 10^{-14}\times f_0= 4.1$~Hz and the line broadening induced is $\pm 2.3 \times 10^{-16} \times f_0=\pm
0.09$~Hz for the ring radius of 20 $\mu$m. For the ring radius of 40 $\mu$m, the shift $\delta f_S^0 (rf)$ is
$4.0 \times 10^{-14} \times f_0= 16.5$~Hz and the induced broadening is $\pm 4.6 \times 10^{-16} \times f_0 =\pm
0.19$~Hz.  The fractional long
term instability induced by $\Delta N=- 1$ is $5.8 \times 10^{-17}$
for the 10 ion ring of radius 20 $\mu$m (and $3.5 \times 10^{-17}$
for the 20 ion ring of radius 40 $\mu$m).

The   tensorial Stark shift  induced by the rf electric field
depends on  the angle $\theta$ between the local electric field and
the magnetic field like \cite{itano00} 
\beq \delta
f_S^2(rf)=-\frac{1}{2}\Delta \alpha^2 f(M_J) \frac{3 \cos^2 \theta
-1}{2}\frac{E_{rf}^2}{2}.
 \eeq
 In an rf trap, the amplitude of the rf
electric field is constant over a ring but its direction rotates in
the radial plane. The projection of this electric field on a static
frame    is defined by $(\cos(k-1)\alpha, \sin(k-1)\alpha)$ if
$\alpha$ is the angle between the ion radial direction and a
reference axis chosen in the radial plane. If the geometric
configuration is such that $\theta$ varies from one ion to the
other, the dispersion of the Stark shift induces a line broadening
of the order of 1 Hz for the 20 $\mu$m radius ring and of 4 Hz for
the 40 $\mu$m radius ring. Such a broadening, which is bigger than
any other broadening calculated so far, can be avoided by choosing a
magnetic field oriented along the trap axis ($\theta=\pi/2$),
to assure the same Stark shift  for all the ions of the ring. In
this configuration and with the same trapping parameters, the
tensorial Stark effect is $\delta f_S^2 (rf)/f(M_J) = -0.3 \times
10^{-14}  \times f_0= -1.1$~H, with a line broadening $\pm 0.6 \times 10^{-16} \times f_0=\pm 0.02$~Hz,
for the ring radius of 20 $\mu$m ($\delta f_S^2 (rf)/f(M_J) =-1.1
\times 10^{-14}  \times f_0= -4.6$~Hz with a broadening of $\pm 1.3 \times 10^{-16}\times f_0= \pm
0.05$~Hz for  the ring radius of 40 $\mu$m). For the first case,
the fractional long term frequency fluctuation induced by the loss of one ion
is $1.4 \times 10^{-17}\times f(M_J)$ and is in the $10^{-18}$
range for the larger ring. The choice of the Zeeman sublevel used
for the clock operation results from a compromise between several
shifts, including the Zeeman shift itself. In the next section we
present these effects that are not directly related to the trapping
electric field.

\section{Other systematic shifts and their dispersion}\label{s_others}
\subsection{ Zeeman effect}\label{s_zeeman}
The selection rules for a quadrupole transition forbid $\Delta
M_J=0$  transition when the magnetic field lies along the direction
of propagation of the clock laser. As the magnetic field is chosen
along the trap symmetry axis to prevent the dispersion of the
tensorial Stark shift, the use of $\Delta M_J=0$ transitions
would require an angle between the trap axis and the direction of
propagation of the laser. This option has to be rejected because of
the large Doppler effect induced by the rf driven motion. As
a second choice, the $\Delta M_J=\pm 1$ transitions are allowed
when the laser propagates along the magnetic field direction.
Indeed, the (S$_{1/2}, M_J= \pm 1/2 \rightarrow$ D$_{5/2}, M_J= \mp
1/2$) transitions are only four times more sensitive to magnetic
field fluctuations than their (S$_{1/2}, M_J= \pm 1/2 \rightarrow$
D$_{5/2}, M_J= \pm 1/2$) counterparts. The summation of the
frequency of the two transitions (S$_{1/2}, M_J= \pm 1/2
\rightarrow$ D$_{5/2}, M_J= \mp 1/2$) should cancel the first order
Zeeman shift if the magnetic field is constant in time at every
point of the ring. A field fluctuation $\delta B$ induces an
uncertainty and a line broadening on this transition of 2.2 MHz/G.
From the frequency fluctuations evaluated in   \cite{burt02} for the
JPL Hg$^+$ microwave clock, and reached by a three layer magnetic
shielding, one can infer magnetic field fluctuations of the order of
$6 \times 10^{-7}$ G, for a magnetic field of 0.05 G.

The magnetic field  has to be  large enough to resolve the
transitions between various Zeeman sublevels. In a Ca$^+$
experiment, having a 1 kHz  separation between neighboring
transitions requires a magnetic field of $6 \times 10^{-4}$ G. If
one assumes the relative stability demonstrated in the JPL microwave
clock, the Zeeman shift fluctuations are of the order of 0.013 Hz
which result in a fractional uncertainty of $3 \times 10^{-17}$. If
the same absolute stability can be reproduced, the Zeeman shift
fluctuations reach 1.0 Hz which is equivalent to a fractional
uncertainty of $2.5 \times 10^{-15}$. The same degree of frequency
stability can be reached with less stringent conditions on the
magnetic field stability using the hyperfine transition $M_F=0 \to
M_F^{'}=0$ of the odd isotope $^{43}$Ca$^+$ \cite{champenois04,
kajita05}. Nevertheless, cooling and state detection with an odd
isotope require more laser sources than with an even isotope and
this complexity may not be acceptable.

The Zeeman effect  also reduces the stability of a single ion cloud.
The main difference here is that the magnetic field has to be kept
constant and homogeneous over the 40 $\mu$m (or 80 $\mu$m) of the
ring diameter, compared to the 1 $\mu$m scale associated to a single
ion. This is still a favorable condition compared to the several centimeters long cloud
used at JPL.

\subsection{Blackbody radiation shift}\label{s_bbr}
The blackbody radiation shift (BBR shift) is the Stark effect
induced by the thermal electric field radiated by the vacuum vessel
and every part inside it.  For calcium  ions,  very precise
theoretical calculations  \cite{arora07} for the polarizabilities
implied in the BBR shift allow to know the shift to better than 3\%:
$\delta f_{BBR}=0.38(1)$ Hz. This result assumes that the radiated
field is isotropic and that the temperature of the emitting surface
is 300 K. As the BBR shift scales with $T^4$, thermal fluctuations
can be detrimental to high precision clocks. A 10~K uncertainty
keeps the frequency uncertainty and eventual broadening (less
probabe as the time scales involved are very long) at the 0.05~Hz
level, which is negligible in the context of our clock project. In
conclusion, for temperature fluctuations smaller than 10~K a long
term instability lower than $10^{-16}$ can be reached.

\subsection{Quadrupole shift}\label{s_quad}
Another effect well known in single ion optical clocks is the
quadrupole shift induced by the gradient of the local electric field
coupled to the electric quadrupole of the D$_{5/2}$ state (a
S$_{1/2}$ state has none). Thanks to a recent experiment
\cite{roos06},  in agreement with precise theoretical calculations
\cite{jiang08}, the quadrupole of the Ca$^+$ D$_{5/2}$ state is
known to better than 1\%: $\theta($D$_{5/2}$)=1.83(1) $e a_0^2$
(atomic unit). The coupling with the rf electric field gradient
gives rise to sidebands in the clock transition spectra that are
well separated from the central band $f_0$, so that the dc  electric
field gradient is the only relevant contribution for this shift. The
dc trapping field has a quadrupole profile and using the notation of
\cite{itano00}, its  gradient is $2A=-m\omega_z^2/(2q)$. For a
magnetic field parallel to the trap symmetry axis and for
$\omega_z/2\pi=1$~MHz, the quadrupole shift is \beq \delta f_Q=1.0
\times (3M_J^2-35/4) {\rm  \ \ Hz}. \eeq For the Zeeman sublevels
$M_J=\pm 1/2$, the quadrupole shift induced by this dc trapping
field is 8.0 Hz and does not depend on the position of the ions. If
the axial trapping voltage is well controlled,   this shift should
not reduce the performance of the clock. It is not the case for the
shift induced by extra dc fields which can build up in the trap,
for example by neutral atom deposition. These fields can easily
induce shifts of the order of 1 Hz from day to day operation and are
well known from single ion clock operation. Two methods are used to
reduce their day to day fluctuations: ionizing the neutral beam far
from the clock operation area before shuttling them and/or using
photoionization \cite{kjaergaard00,rotter01}  which requires a far
smaller flux of neutrals than the traditional electron bombardment
method. If required, the shift itself can be compensated by the
combination of three measurements on different Zeeman sublevels
\cite{dube05}. The method based on frequency measurements with three
orthogonal directions of the magnetic field, used for single ion
standard \cite{itano00,margolis04}, can not be applied here because
of the tensorial Stark shift dispersion (see \ref{s_stark}).  All
together, the    uncertainty and fluctuations induced by this
quadrupole shift must be made smaller than 0.04 Hz to keep the long
term fractional frequency fluctuations smaller than $10^{-16}$ .

\subsection{Misalignment}\label{s_power}
If the propagation axis of the clock laser does not match  the
symmetry axis of the ring, the ions do not see the same laser power.
The induced variations in the excitation probability contribute to
the noise of the excitation detection and reduce the short term
stability of the clock. Assuming a single Rabi pulse to interrogate
the clock transition and a maximum excitation probability on
resonance ($\Omega_L T=\pi$ with $\Omega_L$ the Rabi frequency of
the atom-laser coupling  and $T$ the pulse duration), the variation
in the excitation probability for a detuning giving the maximum
sensitivity is $\delta P_e(T)=0.72 \delta \Omega_L/\Omega_L$. If one
considers a Gaussian laser beam of waist $w_L$ and intensity profile
$I_L(r)=I_0 \exp(-2r^2/w_L^2)$, one can connect the variation
$\delta \Omega_L$ to a misalignment $\delta r$. The dispersion in
the probability of excitation is then 
\beq
 \delta P_e(T)=0.36
\frac{4 r \delta r}{w_L^2}.
 \eeq 
The quantum projection noise
$1/2 \sqrt{N}$ \cite{itano93} is a fundamental limit which can only
be beaten  by entanglement and squeezing methods  \cite{wineland94}.
Making $\delta P_e(T)$ small compared to this noise is sufficient to
make misalignment effects negligible. An easy solution is to make
the laser waist a lot larger than the ring radius $R$. Already with
$w_L=2 R$, a geometrical precision of $\delta r < 0.27 R$ is enough
to keep $ \delta P_e(T) <0.1$ and makes the contribution to the
noise negligible.

\section{Conclusion}\label{s_fin}
The shift and uncertainty budget presented in table
\ref{tab_precision} shows that  the rf is responsible for the major
shifts and broadenings. The total broadening of the transition  is
smaller than 1 Hz, which is smaller than or comparable to the
spectral broadening induced by the finite time excitation on the
clock transition, depending on the Rabi pulse duration
\cite{champenois04}. Therefore, the broadening due to the rf induced
shifts does  alter the short term stability and having 10 or 20 ions
instead of one effectively results in a gain in stability by
$\sqrt{N}$.
\begin{table}
\caption{Uncertainty budget for the frequency transition of $\left|S_{1/2}, M_J=\pm 1/2 \right> \rightarrow \left|D_{5/2},M_J=\mp 1/2 \right>$ in $^{40}$Ca$^{+}$, based on a ring in an octupole linear trap with $\omega_z/2\pi=1$  MHz, $\Omega/2\pi$=20 MHz,  and a rf electric field such that $R=20 \mu$m (10 ions) or $40 \mu$m (20 ions), as given in the table.} \label{tab_precision}
\begin{center}
\begin{tabular}{|c|c|c|c|c|}
\hline
effect & conditions &  shift (Hz) & broadening &long term instability \\
\hline \hline
 Doppler($2^e$) &  $R=20 \mu$m & $ +6.0 $& $\pm 0.14$ & $8 \times 10^{-17}$  \\
    &  $R=40 \mu$m & $ +24.1 $& $\pm 0.28$ & $5 \times 10^{-17}$  \\
 Stark  &   $R=20 \mu$m  & $+4.1$  &  $\pm 0.09$ & $6 \times 10^{-17}$ \\
  &  $R=40 \mu$m  & $+16.5$  &  $\pm 0.19$ & $3 \times 10^{-17}$ \\
 Zeeman  &  $\delta B \le 6 \times 10^{-7}$  G &  & $<1$ & $ 2.5 \times 10^{-15}$ \\
BBR & $T=300 \pm 10$ K  & $+0.38(1) \pm 0.05$  &  &$ <10^{-16}$ \\
quadrupole & trapping field & +8.0  &$<0.1$ &   $\leq 10^{-17}$\\
quadrupole & extra dc & &  $\simeq 0.04$&   $\leq 10^{-16}$\\
\hline
total & $R=20 \mu$m    & +18.5  &  $\pm 0.2$  &  $ 2.5 \times 10^{-15}$ \\
\hline
total & $R=40 \mu$m    & +49.0 &  $\pm 0.4$ &   $ 2.5 \times 10^{-15}$ \\
\hline

\end{tabular}
\end{center}
\end{table}

Furthermore, the fluctuations of the rf induced effects over long
time scales are not the limiting factors for the long term stability
of the clock. Indeed, a single ion standard encounters the same
limitations concerning uncertainty and long term stability, due to
coupling to the local electric and magnetic fields. The challenge
for a ring is to keep the stability constrains over larger spatial
scales than for a single ion.

As we have seen earlier, increasing  the number of trapped ions
allows to further reduce the long term fluctuations induced by  ion
loss, which scale like $\delta N/N^3$ and to lower the short term
instability by increasing the signal to noise ratio. In this
article, we limited our study to 10 and 20 ions ring to keep the
temperature compatible with clock operation conditions while the
cooling lasers are off.  An alternative to a sequential operation
that could allow to work with larger rings is to sympathetically
cool the calcium ions by ions with a different mass (LC ions). Our
MD simulations on 10 calcium ions show that conditions for axial and
radial decoupling  can be found, allowing the LC ions to reach their
Doppler limit on the axial motion while calcium axial temperature
fluctuates between 1 and 3 mK. These studies have to be extended
\cite{marciante_tocome} to larger samples to demonstrate a gain
compared to the simple one species ring we have considered in this
article. Such large cold rings could find an interest in metrology
and/or quantum information or simulation.  With today's state of the
art in magnetic field stabilisation, the ion loss is not
the limiting effect for the frequency stability of a clock based on
a ring of ions and this configuration can offer the possibility to
test many particles system for metrology.

\begin{acknowledgments} One of the authors (C.C.) thanks Tanja Mehlst\"aubler for stimulating discussions about the ring configuration. Fernande Vedel is gratefully acknowledged for her suggestions. M.~Kajita  was supported by a visiting professor grant at the
Universit\'e de Provence when part of this work was initiated. \end{acknowledgments}


\begin{thebibliography}{42}
\expandafter\ifx\csname natexlab\endcsname\relax\def\natexlab#1{#1}\fi
\expandafter\ifx\csname bibnamefont\endcsname\relax
  \def\bibnamefont#1{#1}\fi
\expandafter\ifx\csname bibfnamefont\endcsname\relax
  \def\bibfnamefont#1{#1}\fi
\expandafter\ifx\csname citenamefont\endcsname\relax
  \def\citenamefont#1{#1}\fi
\expandafter\ifx\csname url\endcsname\relax
  \def\url#1{\texttt{#1}}\fi
\expandafter\ifx\csname urlprefix\endcsname\relax\def\urlprefix{URL }\fi
\providecommand{\bibinfo}[2]{#2}
\providecommand{\eprint}[2][]{\url{#2}}

\bibitem[{\citenamefont{Udem et~al.}(2001)\citenamefont{Udem, Diddams, Vogel,
  Oates, Curtis, Lee, Itano, Drullinger, Bergquist, and Hollberg}}]{udem01}
\bibinfo{author}{\bibfnamefont{T.}~\bibnamefont{Udem}},
  \bibinfo{author}{\bibfnamefont{S.~A.} \bibnamefont{Diddams}},
  \bibinfo{author}{\bibfnamefont{K.~R.} \bibnamefont{Vogel}},
  \bibinfo{author}{\bibfnamefont{C.~W.} \bibnamefont{Oates}},
  \bibinfo{author}{\bibfnamefont{E.~A.} \bibnamefont{Curtis}},
  \bibinfo{author}{\bibfnamefont{W.~D.} \bibnamefont{Lee}},
  \bibinfo{author}{\bibfnamefont{W.~M.} \bibnamefont{Itano}},
  \bibinfo{author}{\bibfnamefont{R.~E.} \bibnamefont{Drullinger}},
  \bibinfo{author}{\bibfnamefont{J.~C.} \bibnamefont{Bergquist}},
  \bibnamefont{and} \bibinfo{author}{\bibfnamefont{L.}~\bibnamefont{Hollberg}},
  \bibinfo{journal}{Phys. Rev. Lett.} \textbf{\bibinfo{volume}{86}},
  \bibinfo{pages}{4996} (\bibinfo{year}{2001}).

\bibitem[{\citenamefont{Degenhardt et~al.}(2005)\citenamefont{Degenhardt,
  Stoehr, Lisdat, Wilpers, Schnatz, Lipphardt, Nazarova, Pottie, Sterr, Helmcke
  et~al.}}]{degenhardt05}
\bibinfo{author}{\bibfnamefont{C.}~\bibnamefont{Degenhardt}},
  \bibinfo{author}{\bibfnamefont{H.}~\bibnamefont{Stoehr}},
  \bibinfo{author}{\bibfnamefont{C.}~\bibnamefont{Lisdat}},
  \bibinfo{author}{\bibfnamefont{G.}~\bibnamefont{Wilpers}},
  \bibinfo{author}{\bibfnamefont{H.}~\bibnamefont{Schnatz}},
  \bibinfo{author}{\bibfnamefont{B.}~\bibnamefont{Lipphardt}},
  \bibinfo{author}{\bibfnamefont{T.}~\bibnamefont{Nazarova}},
  \bibinfo{author}{\bibfnamefont{P.-E.} \bibnamefont{Pottie}},
  \bibinfo{author}{\bibfnamefont{U.}~\bibnamefont{Sterr}},
  \bibinfo{author}{\bibfnamefont{J.}~\bibnamefont{Helmcke}},
  \bibnamefont{et~al.}, \bibinfo{journal}{Phys. Rev. A}
  \textbf{\bibinfo{volume}{72}}, \bibinfo{pages}{062111}
  (\bibinfo{year}{2005}).

\bibitem[{\citenamefont{Takamoto et~al.}(2005)\citenamefont{Takamoto, Hong,
  Higashi, and Katori}}]{takamoto05}
\bibinfo{author}{\bibfnamefont{M.}~\bibnamefont{Takamoto}},
  \bibinfo{author}{\bibfnamefont{F.-L.} \bibnamefont{Hong}},
  \bibinfo{author}{\bibfnamefont{R.}~\bibnamefont{Higashi}}, \bibnamefont{and}
  \bibinfo{author}{\bibfnamefont{H.}~\bibnamefont{Katori}},
  \bibinfo{journal}{Nature} \textbf{\bibinfo{volume}{435}},
  \bibinfo{pages}{321} (\bibinfo{year}{2005}).

\bibitem[{\citenamefont{Akatsuka et~al.}(2008)\citenamefont{Akatsuka, Takamoto,
  and Katori}}]{akatsuka08}
\bibinfo{author}{\bibfnamefont{T.}~\bibnamefont{Akatsuka}},
  \bibinfo{author}{\bibfnamefont{M.}~\bibnamefont{Takamoto}}, \bibnamefont{and}
  \bibinfo{author}{\bibfnamefont{H.}~\bibnamefont{Katori}},
  \bibinfo{journal}{Nature Physics} \textbf{\bibinfo{volume}{4}},
  \bibinfo{pages}{954} (\bibinfo{year}{2008}).

\bibitem[{\citenamefont{Lemke et~al.}(2009)\citenamefont{Lemke, Ludlow, Barber,
  Fortier, Diddams, Jiang, Jefferts, Heavner, Parker, and Oates}}]{lemke09}
\bibinfo{author}{\bibfnamefont{N.~D.} \bibnamefont{Lemke}},
  \bibinfo{author}{\bibfnamefont{A.~D.} \bibnamefont{Ludlow}},
  \bibinfo{author}{\bibfnamefont{Z.~W.} \bibnamefont{Barber}},
  \bibinfo{author}{\bibfnamefont{T.~M.} \bibnamefont{Fortier}},
  \bibinfo{author}{\bibfnamefont{S.~A.} \bibnamefont{Diddams}},
  \bibinfo{author}{\bibfnamefont{Y.}~\bibnamefont{Jiang}},
  \bibinfo{author}{\bibfnamefont{S.~R.} \bibnamefont{Jefferts}},
  \bibinfo{author}{\bibfnamefont{T.~P.} \bibnamefont{Heavner}},
  \bibinfo{author}{\bibfnamefont{T.~E.} \bibnamefont{Parker}},
  \bibnamefont{and} \bibinfo{author}{\bibfnamefont{C.~W.} \bibnamefont{Oates}},
  \bibinfo{journal}{Phys. Rev. Lett.} \textbf{\bibinfo{volume}{103}},
  \bibinfo{pages}{063001} (\bibinfo{year}{2009}).

\bibitem[{\citenamefont{Margolis et~al.}(2004)\citenamefont{Margolis, Barwood,
  Huang, Klein, Lea, Szymaniec, and Gill}}]{margolis04}
\bibinfo{author}{\bibfnamefont{H.~S.} \bibnamefont{Margolis}},
  \bibinfo{author}{\bibfnamefont{G.~P.} \bibnamefont{Barwood}},
  \bibinfo{author}{\bibfnamefont{G.}~\bibnamefont{Huang}},
  \bibinfo{author}{\bibfnamefont{H.~A.} \bibnamefont{Klein}},
  \bibinfo{author}{\bibfnamefont{S.~N.} \bibnamefont{Lea}},
  \bibinfo{author}{\bibfnamefont{K.}~\bibnamefont{Szymaniec}},
  \bibnamefont{and} \bibinfo{author}{\bibfnamefont{P.}~\bibnamefont{Gill}},
  \bibinfo{journal}{Science} \textbf{\bibinfo{volume}{306}},
  \bibinfo{pages}{1355} (\bibinfo{year}{2004}).

\bibitem[{\citenamefont{Madej et~al.}(2004)\citenamefont{Madej, Bernard,
  Dub\'e, Marmet, and Windeler}}]{madej04}
\bibinfo{author}{\bibfnamefont{A.~A.} \bibnamefont{Madej}},
  \bibinfo{author}{\bibfnamefont{J.~E.} \bibnamefont{Bernard}},
  \bibinfo{author}{\bibfnamefont{P.}~\bibnamefont{Dub\'e}},
  \bibinfo{author}{\bibfnamefont{L.}~\bibnamefont{Marmet}}, \bibnamefont{and}
  \bibinfo{author}{\bibfnamefont{R.~S.} \bibnamefont{Windeler}},
  \bibinfo{journal}{Phys. Rev.~A} \textbf{\bibinfo{volume}{70}},
  \bibinfo{eid}{012507} (\bibinfo{year}{2004}).

\bibitem[{\citenamefont{Schneider et~al.}(2005)\citenamefont{Schneider, Peik,
  and Tamm}}]{schneider05}
\bibinfo{author}{\bibfnamefont{T.}~\bibnamefont{Schneider}},
  \bibinfo{author}{\bibfnamefont{E.}~\bibnamefont{Peik}}, \bibnamefont{and}
  \bibinfo{author}{\bibfnamefont{C.}~\bibnamefont{Tamm}},
  \bibinfo{journal}{Phys. Rev. Lett.} \textbf{\bibinfo{volume}{94}},
  \bibinfo{eid}{230801} (\bibinfo{year}{2005}).

\bibitem[{\citenamefont{Rosenband et~al.}(2008)\citenamefont{Rosenband, Hume,
  Schmidt, Chou, Brusch, Lorini, Oskay, Drullinger, Fortier, Stalnaker
  et~al.}}]{rosenband08}
\bibinfo{author}{\bibfnamefont{T.}~\bibnamefont{Rosenband}},
  \bibinfo{author}{\bibfnamefont{D.~B.} \bibnamefont{Hume}},
  \bibinfo{author}{\bibfnamefont{P.~O.} \bibnamefont{Schmidt}},
  \bibinfo{author}{\bibfnamefont{C.~W.} \bibnamefont{Chou}},
  \bibinfo{author}{\bibfnamefont{A.}~\bibnamefont{Brusch}},
  \bibinfo{author}{\bibfnamefont{L.}~\bibnamefont{Lorini}},
  \bibinfo{author}{\bibfnamefont{W.~H.} \bibnamefont{Oskay}},
  \bibinfo{author}{\bibfnamefont{R.~E.} \bibnamefont{Drullinger}},
  \bibinfo{author}{\bibfnamefont{T.~M.} \bibnamefont{Fortier}},
  \bibinfo{author}{\bibfnamefont{J.~E.} \bibnamefont{Stalnaker}},
  \bibnamefont{et~al.}, \bibinfo{journal}{Science}
  \textbf{\bibinfo{volume}{319}}, \bibinfo{pages}{1808} (\bibinfo{year}{2008}).

\bibitem[{\citenamefont{Fisk}(1997)}]{fisk97}
\bibinfo{author}{\bibfnamefont{P.}~\bibnamefont{Fisk}}, \bibinfo{journal}{Rep.
  Prog. Phys.} \textbf{\bibinfo{volume}{60}}, \bibinfo{pages}{761}
  (\bibinfo{year}{1997}).

\bibitem[{\citenamefont{Prestage et~al.}(2006)\citenamefont{Prestage, Chung,
  Le, Lim, and Maleki}}]{prestage06}
\bibinfo{author}{\bibfnamefont{J.}~\bibnamefont{Prestage}},
  \bibinfo{author}{\bibfnamefont{S.}~\bibnamefont{Chung}},
  \bibinfo{author}{\bibfnamefont{T.}~\bibnamefont{Le}},
  \bibinfo{author}{\bibfnamefont{L.}~\bibnamefont{Lim}}, \bibnamefont{and}
  \bibinfo{author}{\bibfnamefont{L.}~\bibnamefont{Maleki}}, in
  \emph{\bibinfo{booktitle}{Proceedings of IEEE Int. Freq. Contr. Symp. Miami,
  jun. 5-7, 2006}} (\bibinfo{publisher}{IEEE, New York}, \bibinfo{year}{2006}).

\bibitem[{\citenamefont{Prestage and Weaver}(2007)}]{prestage07}
\bibinfo{author}{\bibfnamefont{J.}~\bibnamefont{Prestage}} \bibnamefont{and}
  \bibinfo{author}{\bibfnamefont{G.}~\bibnamefont{Weaver}},
  \bibinfo{journal}{Proceeding of the IEEE} \textbf{\bibinfo{volume}{95}},
  \bibinfo{pages}{2235} (\bibinfo{year}{2007}).

\bibitem[{\citenamefont{Prestage et~al.}(1999)\citenamefont{Prestage, Tjoelker,
  and Maleki}}]{prestage99}
\bibinfo{author}{\bibfnamefont{J.}~\bibnamefont{Prestage}},
  \bibinfo{author}{\bibfnamefont{R.}~\bibnamefont{Tjoelker}}, \bibnamefont{and}
  \bibinfo{author}{\bibfnamefont{L.}~\bibnamefont{Maleki}},
  \bibinfo{journal}{Proceedings of the 1999 Joint EFTF-IFCS, Besancon, France}
  pp. \bibinfo{pages}{121--124} (\bibinfo{year}{1999}).

\bibitem[{\citenamefont{Prestage et~al.}(2001)\citenamefont{Prestage, Tjoelker,
  and Maleki}}]{prestage01}
\bibinfo{author}{\bibfnamefont{J.}~\bibnamefont{Prestage}},
  \bibinfo{author}{\bibfnamefont{R.}~\bibnamefont{Tjoelker}}, \bibnamefont{and}
  \bibinfo{author}{\bibfnamefont{L.}~\bibnamefont{Maleki}},
  \emph{\bibinfo{title}{Frequency measurement and control: Advanced techniques
  and future trends}} (\bibinfo{publisher}{Springer, Berlin},
  \bibinfo{year}{2001}), chap. \bibinfo{chapter}{Recent developments in
  microwave ion clocks}.

\bibitem[{\citenamefont{Gerlich}(1992)}]{gerlich92}
\bibinfo{author}{\bibfnamefont{D.}~\bibnamefont{Gerlich}}, in
  \emph{\bibinfo{booktitle}{State-selected and state-to-state ion-molecule
  reaction dynamics, Part I}}, edited by \bibinfo{editor}{\bibfnamefont{C.-Y.}
  \bibnamefont{Ng}} \bibnamefont{and}
  \bibinfo{editor}{\bibfnamefont{M.}~\bibnamefont{Baer}}
  (\bibinfo{publisher}{John Wiley and Sons}, \bibinfo{year}{1992}),
  vol.~\bibinfo{volume}{82} of \emph{\bibinfo{series}{Advances in Chemical
  Physics Series}}.

\bibitem[{\citenamefont{Wester}(2009)}]{wester09}
\bibinfo{author}{\bibfnamefont{R.}~\bibnamefont{Wester}}, \bibinfo{journal}{J.
  Phys. B} \textbf{\bibinfo{volume}{42}}, \bibinfo{pages}{154001}
  (\bibinfo{year}{2009}).

\bibitem[{\citenamefont{Okada et~al.}(2007)\citenamefont{Okada, Yasuda,
  Takayanagi, Wada, Schuessler, and Ohtani}}]{okada07}
\bibinfo{author}{\bibfnamefont{K.}~\bibnamefont{Okada}},
  \bibinfo{author}{\bibfnamefont{K.}~\bibnamefont{Yasuda}},
  \bibinfo{author}{\bibfnamefont{T.}~\bibnamefont{Takayanagi}},
  \bibinfo{author}{\bibfnamefont{M.}~\bibnamefont{Wada}},
  \bibinfo{author}{\bibfnamefont{H.~A.} \bibnamefont{Schuessler}},
  \bibnamefont{and} \bibinfo{author}{\bibfnamefont{S.}~\bibnamefont{Ohtani}},
  \bibinfo{journal}{Phys. Rev. A} \textbf{\bibinfo{volume}{75}},
  \bibinfo{eid}{033409} (\bibinfo{year}{2007}).

\bibitem[{\citenamefont{Okada et~al.}(2009)\citenamefont{Okada, Takayanagi,
  Wada, Ohtani, and Schuessler}}]{okada09}
\bibinfo{author}{\bibfnamefont{K.}~\bibnamefont{Okada}},
  \bibinfo{author}{\bibfnamefont{T.}~\bibnamefont{Takayanagi}},
  \bibinfo{author}{\bibfnamefont{M.}~\bibnamefont{Wada}},
  \bibinfo{author}{\bibfnamefont{S.}~\bibnamefont{Ohtani}}, \bibnamefont{and}
  \bibinfo{author}{\bibfnamefont{H.~A.} \bibnamefont{Schuessler}},
  \bibinfo{journal}{Phys. Rev.~A} \textbf{\bibinfo{volume}{80}},
  \bibinfo{pages}{043405} (\bibinfo{year}{2009}).

\bibitem[{\citenamefont{Calvo et~al.}(2009)\citenamefont{Calvo, Champenois, and
  Yurtsever}}]{calvo09}
\bibinfo{author}{\bibfnamefont{F.}~\bibnamefont{Calvo}},
  \bibinfo{author}{\bibfnamefont{C.}~\bibnamefont{Champenois}},
  \bibnamefont{and}
  \bibinfo{author}{\bibfnamefont{E.}~\bibnamefont{Yurtsever}},
  \bibinfo{journal}{Phys. Rev.~A} \textbf{\bibinfo{volume}{80}},
  \bibinfo{eid}{063401} (pages~\bibinfo{numpages}{6}) (\bibinfo{year}{2009}).

\bibitem[{\citenamefont{Champenois}(2009)}]{champenois09}
\bibinfo{author}{\bibfnamefont{C.}~\bibnamefont{Champenois}},
  \bibinfo{journal}{J. Phys. B} \textbf{\bibinfo{volume}{42}},
  \bibinfo{pages}{154002} (\bibinfo{year}{2009}).

\bibitem[{\citenamefont{Itano et~al.}(1993)\citenamefont{Itano, Bergquist,
  Bollinger, Gilligan, Heinzen, Moore, Raizen, and Wineland}}]{itano93}
\bibinfo{author}{\bibfnamefont{W.}~\bibnamefont{Itano}},
  \bibinfo{author}{\bibfnamefont{J.}~\bibnamefont{Bergquist}},
  \bibinfo{author}{\bibfnamefont{J.}~\bibnamefont{Bollinger}},
  \bibinfo{author}{\bibfnamefont{J.}~\bibnamefont{Gilligan}},
  \bibinfo{author}{\bibfnamefont{D.}~\bibnamefont{Heinzen}},
  \bibinfo{author}{\bibfnamefont{F.}~\bibnamefont{Moore}},
  \bibinfo{author}{\bibfnamefont{M.}~\bibnamefont{Raizen}}, \bibnamefont{and}
  \bibinfo{author}{\bibfnamefont{D.}~\bibnamefont{Wineland}},
  \bibinfo{journal}{Phys. Rev.~A} \textbf{\bibinfo{volume}{47}},
  \bibinfo{pages}{3554} (\bibinfo{year}{1993}).

\bibitem[{\citenamefont{Porras and Cirac}(2006)}]{porras06}
\bibinfo{author}{\bibfnamefont{D.}~\bibnamefont{Porras}} \bibnamefont{and}
  \bibinfo{author}{\bibfnamefont{J.~I.} \bibnamefont{Cirac}},
  \bibinfo{journal}{Phys. Rev. Lett.} \textbf{\bibinfo{volume}{96}},
  \bibinfo{pages}{250501} (\bibinfo{year}{2006}).

\bibitem[{\citenamefont{Buluta et~al.}(2008)\citenamefont{Buluta, Kitaoka,
  Georgescu, and Hasegawa}}]{buluta08}
\bibinfo{author}{\bibfnamefont{I.~M.} \bibnamefont{Buluta}},
  \bibinfo{author}{\bibfnamefont{M.}~\bibnamefont{Kitaoka}},
  \bibinfo{author}{\bibfnamefont{S.}~\bibnamefont{Georgescu}},
  \bibnamefont{and} \bibinfo{author}{\bibfnamefont{S.}~\bibnamefont{Hasegawa}},
  \bibinfo{journal}{Phys. Rev.~A} \textbf{\bibinfo{volume}{77}},
  \bibinfo{pages}{062320} (\bibinfo{year}{2008}).

\bibitem[{\citenamefont{Olmos et~al.}(2009)\citenamefont{Olmos,
  Gonz\'alez-F\'erez, and Lesanovsky}}]{olmos09}
\bibinfo{author}{\bibfnamefont{B.}~\bibnamefont{Olmos}},
  \bibinfo{author}{\bibfnamefont{R.}~\bibnamefont{Gonz\'alez-F\'erez}},
  \bibnamefont{and}
  \bibinfo{author}{\bibfnamefont{I.}~\bibnamefont{Lesanovsky}},
  \bibinfo{journal}{Phys. Rev. Lett.} \textbf{\bibinfo{volume}{103}},
  \bibinfo{pages}{185302} (\bibinfo{year}{2009}).

\bibitem[{\citenamefont{Champenois et~al.}(2004)\citenamefont{Champenois,
  Houssin, Lisowski, Knoop, Vedel, and Vedel}}]{champenois04}
\bibinfo{author}{\bibfnamefont{C.}~\bibnamefont{Champenois}},
  \bibinfo{author}{\bibfnamefont{M.}~\bibnamefont{Houssin}},
  \bibinfo{author}{\bibfnamefont{C.}~\bibnamefont{Lisowski}},
  \bibinfo{author}{\bibfnamefont{M.}~\bibnamefont{Knoop}},
  \bibinfo{author}{\bibfnamefont{M.}~\bibnamefont{Vedel}}, \bibnamefont{and}
  \bibinfo{author}{\bibfnamefont{F.}~\bibnamefont{Vedel}},
  \bibinfo{journal}{Phys. Lett. A} \textbf{\bibinfo{volume}{331}},
  \bibinfo{pages}{298} (\bibinfo{year}{2004}).

\bibitem[{\citenamefont{Kajita et~al.}(2005)\citenamefont{Kajita, Li,
  Matsubara, Hayasaka, and Hosokawa}}]{kajita05}
\bibinfo{author}{\bibfnamefont{M.}~\bibnamefont{Kajita}},
  \bibinfo{author}{\bibfnamefont{Y.}~\bibnamefont{Li}},
  \bibinfo{author}{\bibfnamefont{K.}~\bibnamefont{Matsubara}},
  \bibinfo{author}{\bibfnamefont{K.}~\bibnamefont{Hayasaka}}, \bibnamefont{and}
  \bibinfo{author}{\bibfnamefont{M.}~\bibnamefont{Hosokawa}},
  \bibinfo{journal}{Phys. Rev.~A} \textbf{\bibinfo{volume}{72}},
  \bibinfo{pages}{043404} (\bibinfo{year}{2005}).

\bibitem[{\citenamefont{Teloy and Gerlich}(1974)}]{teloy74}
\bibinfo{author}{\bibfnamefont{E.}~\bibnamefont{Teloy}} \bibnamefont{and}
  \bibinfo{author}{\bibfnamefont{D.}~\bibnamefont{Gerlich}},
  \bibinfo{journal}{Chemical Physics} \textbf{\bibinfo{volume}{4}},
  \bibinfo{pages}{417 } (\bibinfo{year}{1974}).

\bibitem[{\citenamefont{Mikosch et~al.}(2008)\citenamefont{Mikosch,
  Fr\"{u}hling, Trippel, Otto, Hlavenka, Schwalm, Weidem\"{u}ller, and
  Wester}}]{mikosch08}
\bibinfo{author}{\bibfnamefont{J.}~\bibnamefont{Mikosch}},
  \bibinfo{author}{\bibfnamefont{U.}~\bibnamefont{Fr\"{u}hling}},
  \bibinfo{author}{\bibfnamefont{S.}~\bibnamefont{Trippel}},
  \bibinfo{author}{\bibfnamefont{R.}~\bibnamefont{Otto}},
  \bibinfo{author}{\bibfnamefont{P.}~\bibnamefont{Hlavenka}},
  \bibinfo{author}{\bibfnamefont{D.}~\bibnamefont{Schwalm}},
  \bibinfo{author}{\bibfnamefont{M.}~\bibnamefont{Weidem\"{u}ller}},
  \bibnamefont{and} \bibinfo{author}{\bibfnamefont{R.}~\bibnamefont{Wester}},
  \bibinfo{journal}{Phys. Rev.~A} \textbf{\bibinfo{volume}{78}},
  \bibinfo{eid}{023402} (pages~\bibinfo{numpages}{13}) (\bibinfo{year}{2008}).

\bibitem[{\citenamefont{Mikosch et~al.}(2007)\citenamefont{Mikosch,
  Fr\"{u}hling, Trippel, Schwalm, Weidem\"{u}ller, and Wester}}]{mikosch07}
\bibinfo{author}{\bibfnamefont{J.}~\bibnamefont{Mikosch}},
  \bibinfo{author}{\bibfnamefont{U.}~\bibnamefont{Fr\"{u}hling}},
  \bibinfo{author}{\bibfnamefont{S.}~\bibnamefont{Trippel}},
  \bibinfo{author}{\bibfnamefont{D.}~\bibnamefont{Schwalm}},
  \bibinfo{author}{\bibfnamefont{M.}~\bibnamefont{Weidem\"{u}ller}},
  \bibnamefont{and} \bibinfo{author}{\bibfnamefont{R.}~\bibnamefont{Wester}},
  \bibinfo{journal}{Phys. Rev. Lett.} \textbf{\bibinfo{volume}{98}},
  \bibinfo{pages}{223001} (\bibinfo{year}{2007}).

\bibitem[{\citenamefont{Raizen et~al.}(1992)\citenamefont{Raizen, Gilligan,
  Bergquist, Itano, and Wineland}}]{raizen92}
\bibinfo{author}{\bibfnamefont{M.~G.} \bibnamefont{Raizen}},
  \bibinfo{author}{\bibfnamefont{J.~M.} \bibnamefont{Gilligan}},
  \bibinfo{author}{\bibfnamefont{J.~C.} \bibnamefont{Bergquist}},
  \bibinfo{author}{\bibfnamefont{W.~M.} \bibnamefont{Itano}}, \bibnamefont{and}
  \bibinfo{author}{\bibfnamefont{D.~J.} \bibnamefont{Wineland}},
  \bibinfo{journal}{Phys. Rev. A} \textbf{\bibinfo{volume}{45}},
  \bibinfo{pages}{6493} (\bibinfo{year}{1992}).

\bibitem[{\citenamefont{Itano et~al.}(1982)\citenamefont{Itano, Lewis, and
  Wineland}}]{itano82}
\bibinfo{author}{\bibfnamefont{W.}~\bibnamefont{Itano}},
  \bibinfo{author}{\bibfnamefont{I.}~\bibnamefont{Lewis}}, \bibnamefont{and}
  \bibinfo{author}{\bibfnamefont{D.}~\bibnamefont{Wineland}},
  \bibinfo{journal}{Phys. Rev.~A} \textbf{\bibinfo{volume}{25}},
  \bibinfo{pages}{1233} (\bibinfo{year}{1982}).

\bibitem[{\citenamefont{Wineland and Itano}(1979)}]{wineland79}
\bibinfo{author}{\bibfnamefont{D.~J.} \bibnamefont{Wineland}} \bibnamefont{and}
  \bibinfo{author}{\bibfnamefont{W.~M.} \bibnamefont{Itano}},
  \bibinfo{journal}{Phys. Rev. A} \textbf{\bibinfo{volume}{20}},
  \bibinfo{pages}{1521} (\bibinfo{year}{1979}).

\bibitem[{\citenamefont{Marciante and {\it et. al}}()}]{marciante_tocome}
\bibinfo{author}{\bibfnamefont{M.}~\bibnamefont{Marciante}} \bibnamefont{and}
  \bibinfo{author}{\bibnamefont{{\it et. al}}}, \bibinfo{howpublished}{in
  preparation}.

\bibitem[{\citenamefont{Itano}(2000)}]{itano00}
\bibinfo{author}{\bibfnamefont{W.}~\bibnamefont{Itano}}, \bibinfo{journal}{J.
  Res. Natl. Inst. Stand. Technol.} \textbf{\bibinfo{volume}{105}},
  \bibinfo{pages}{829} (\bibinfo{year}{2000}).

\bibitem[{\citenamefont{Arora et~al.}(2007)\citenamefont{Arora, Safronova, and
  Clark}}]{arora07}
\bibinfo{author}{\bibfnamefont{B.}~\bibnamefont{Arora}},
  \bibinfo{author}{\bibfnamefont{M.~S.} \bibnamefont{Safronova}},
  \bibnamefont{and} \bibinfo{author}{\bibfnamefont{C.~W.} \bibnamefont{Clark}},
  \bibinfo{journal}{Phys. Rev.~A} \textbf{\bibinfo{volume}{76}},
  \bibinfo{pages}{064501} (\bibinfo{year}{2007}).

\bibitem[{\citenamefont{Burt et~al.}(2002)\citenamefont{Burt, Prestage, and
  Tjoelker}}]{burt02}
\bibinfo{author}{\bibfnamefont{E.~A.} \bibnamefont{Burt}},
  \bibinfo{author}{\bibfnamefont{J.}~\bibnamefont{Prestage}}, \bibnamefont{and}
  \bibinfo{author}{\bibfnamefont{R.}~\bibnamefont{Tjoelker}},
  \bibinfo{journal}{Proceedings of the IEEE IFCS, New Orleans} p.
  \bibinfo{pages}{463} (\bibinfo{year}{2002}).

\bibitem[{\citenamefont{Roos et~al.}(2006)\citenamefont{Roos, Chwalla, Kim,
  Riebe, and Blatt}}]{roos06}
\bibinfo{author}{\bibfnamefont{C.}~\bibnamefont{Roos}},
  \bibinfo{author}{\bibfnamefont{M.}~\bibnamefont{Chwalla}},
  \bibinfo{author}{\bibfnamefont{K.}~\bibnamefont{Kim}},
  \bibinfo{author}{\bibfnamefont{M.}~\bibnamefont{Riebe}}, \bibnamefont{and}
  \bibinfo{author}{\bibfnamefont{R.}~\bibnamefont{Blatt}},
  \bibinfo{journal}{Nature} \textbf{\bibinfo{volume}{443}},
  \bibinfo{pages}{316} (\bibinfo{year}{2006}).

\bibitem[{\citenamefont{Jiang et~al.}(2008)\citenamefont{Jiang, Arora, and
  Safronova}}]{jiang08}
\bibinfo{author}{\bibfnamefont{D.}~\bibnamefont{Jiang}},
  \bibinfo{author}{\bibfnamefont{B.}~\bibnamefont{Arora}}, \bibnamefont{and}
  \bibinfo{author}{\bibfnamefont{M.~S.} \bibnamefont{Safronova}},
  \bibinfo{journal}{Phys. Rev.~A} \textbf{\bibinfo{volume}{78}},
  \bibinfo{pages}{022514} (\bibinfo{year}{2008}).

\bibitem[{\citenamefont{Kj{\ae}rgaard et~al.}(2000)\citenamefont{Kj{\ae}rgaard,
  Hornek{\ae}r, Thommesen, Videsen, and Drewsen}}]{kjaergaard00}
\bibinfo{author}{\bibfnamefont{N.}~\bibnamefont{Kj{\ae}rgaard}},
  \bibinfo{author}{\bibfnamefont{L.}~\bibnamefont{Hornek{\ae}r}},
  \bibinfo{author}{\bibfnamefont{A.~M.} \bibnamefont{Thommesen}},
  \bibinfo{author}{\bibfnamefont{Z.}~\bibnamefont{Videsen}}, \bibnamefont{and}
  \bibinfo{author}{\bibfnamefont{M.}~\bibnamefont{Drewsen}},
  \bibinfo{journal}{Appl. Phys. B} \textbf{\bibinfo{volume}{71}},
  \bibinfo{pages}{207} (\bibinfo{year}{2000}).

\bibitem[{\citenamefont{Gulde et~al.}(2001)\citenamefont{Gulde, Rotter, Barton,
  Schmidt-Kaler, Blatt, and Hogervorst}}]{rotter01}
\bibinfo{author}{\bibfnamefont{S.}~\bibnamefont{Gulde}},
  \bibinfo{author}{\bibfnamefont{D.}~\bibnamefont{Rotter}},
  \bibinfo{author}{\bibfnamefont{P.}~\bibnamefont{Barton}},
  \bibinfo{author}{\bibfnamefont{F.}~\bibnamefont{Schmidt-Kaler}},
  \bibinfo{author}{\bibfnamefont{R.}~\bibnamefont{Blatt}}, \bibnamefont{and}
  \bibinfo{author}{\bibfnamefont{W.}~\bibnamefont{Hogervorst}},
  \bibinfo{journal}{Appl. Phys. B} \textbf{\bibinfo{volume}{73}},
  \bibinfo{pages}{861} (\bibinfo{year}{2001}).

\bibitem[{\citenamefont{Dub\'e et~al.}(2005)\citenamefont{Dub\'e, Madej,
  Bernard, Marmet, Boulanger, and Cundy}}]{dube05}
\bibinfo{author}{\bibfnamefont{P.}~\bibnamefont{Dub\'e}},
  \bibinfo{author}{\bibfnamefont{A.~A.} \bibnamefont{Madej}},
  \bibinfo{author}{\bibfnamefont{J.~E.} \bibnamefont{Bernard}},
  \bibinfo{author}{\bibfnamefont{L.}~\bibnamefont{Marmet}},
  \bibinfo{author}{\bibfnamefont{J.-S.} \bibnamefont{Boulanger}},
  \bibnamefont{and} \bibinfo{author}{\bibfnamefont{S.}~\bibnamefont{Cundy}},
  \bibinfo{journal}{Phys. Rev. Lett.} \textbf{\bibinfo{volume}{95}},
  \bibinfo{eid}{033001} (\bibinfo{year}{2005}).

\bibitem[{\citenamefont{Wineland et~al.}(1994)\citenamefont{Wineland,
  Bollinger, Itano, and Heinzen}}]{wineland94}
\bibinfo{author}{\bibfnamefont{D.}~\bibnamefont{Wineland}},
  \bibinfo{author}{\bibfnamefont{J.}~\bibnamefont{Bollinger}},
  \bibinfo{author}{\bibfnamefont{W.}~\bibnamefont{Itano}}, \bibnamefont{and}
  \bibinfo{author}{\bibfnamefont{D.}~\bibnamefont{Heinzen}},
  \bibinfo{journal}{Phys. Rev.~A} \textbf{\bibinfo{volume}{50}},
  \bibinfo{pages}{67} (\bibinfo{year}{1994}).

\end{thebibliography}

\end{document}